\documentclass[a4paper,aps,twocolumn,floats,showpacs,superscriptaddress]{revtex4}
\usepackage{epsfig}
\newcommand{\avk}{\left< k \right>}
\newcommand{\kin}{k^{\rm in}}
\newcommand{\fluck}{\left< k^2 \right>}

\begin{document}

\title{Topology and correlations in structured scale-free
networks}

\author{Alexei V{\'a}zquez} 
\affiliation{International School for 
Advanced Studies and INFM,
Via Beirut 4, Trieste, I-34014, Italy}
\author{Mari{\'a}n Bogu{\~n}{\'a}} 
\affiliation{Departament de F{\'\i}sica Fonamental, Universitat de
  Barcelona, Av. Diagonal 647, 08028 Barcelona, Spain}
\author {Yamir Moreno} 
\affiliation{The Abdus Salam International Centre for
Theoretical Physics, P.O. Box 586, Trieste,
I-34014, Italy}
\author{Romualdo Pastor-Satorras}  
\affiliation{Departament  de F{\'\i}sica i Enginyeria Nuclear,
  Universitat Polit{\`e}cnica de Catalunya, Campus Nord,
  08034 Barcelona, Spain}
\author{Alessandro Vespignani}
\affiliation{The Abdus Salam International Centre for
Theoretical Physics, P.O. Box 586, Trieste,
I-34014, Italy}

\date{\today}


\begin{abstract}
  We study a recently introduced class of scale-free networks showing
  a high clustering coefficient and non-trivial connectivity
  correlations.  We find that the connectivity probability
  distribution strongly depends on the fine details of the model.  We
  solve exactly the case of low average connectivity, providing also
  exact expressions for the clustering and degree correlation
  functions. The model also exhibits a lack of small world properties
  in the whole parameters range. We discuss the physical properties 
  of these networks in the light of the present detailed analysis. 
\end{abstract}


\maketitle

\section{Introduction}

Recently a major scientific effort has been devoted to the
characterization and modeling of a wide range of social and natural
systems that can be described as networks \cite{barabasi02,dorogorev}.
Systems such as the Internet \cite{falou99,calda00,alexei,alexei02} or
the World-Wide-Web \cite{www99}, social communities \cite{strog01},
food-webs \cite{montoya02}, and biological interacting networks
\cite{wagner01,jeong01,spsk,vazquez} can be represented as a graph
\cite{chartrand}, in which nodes represent the population individuals
and links the physical interactions among them.  Strikingly, many of
these networks have complex topological properties and dynamical
features that cannot be accounted for by classical graph modeling
\cite{erdos60}.  In particular, small-world properties \cite{watts98}
and scale-free degree distributions \cite{barab99} (where the degree
or connectivity of a node is defined as the number of other nodes to
which it is attached) seem to emerge frequently as dominant features
governing the topology of real-world networks.  These global
properties imply a large connectivity heterogeneity and a short
average distance between nodes, which have considerable impact on the
behavior of physical processes taking place on top of the network.
For instance, scale-free (SF) networks have been shown to be resilient
to random damage (absence of a percolation transition)
\cite{barabasi00,newman00,havlin01} and prone to epidemic spreading
(null epidemic threshold) \cite{pv01a,pv01b,lloydsir,moreno02}.

The detailed scrutiny of the topological properties of networks has
pointed out that small-world and scale-free properties come often
along with non-trivial degree correlations and clustering properties.
Recently, an interesting class of networks has been introduced by
Klemm and Egu{\'\i}luz by proposing a growing model in which nodes are
progressively deactivated with a probability inversely proportional to
their connectivity \cite{klemm02}.  Analytical arguments and numerical
simulations have lead to the claim that, under general conditions, the
deactivation model, allowing a core of $m$ active nodes,
generates a network with average degree $\avk=2m$ and degree
probability distribution $P(k)=2m^2 k^{-3}$. Interestingly, the
scale-free properties are associated to a high clustering coefficient.  
For this reason the
deactivation model has been used to study how clustering 
can alter the picture obtained for the
resilience to damage and epidemic spreading in SF networks
\cite{structured,crucitti02}.

In this paper we revisit the analysis of the deactivation model.  We
find an analytical solution in the case of minimal values of 
active nodes $m$ (low average connectivity). In addition, large scale 
numerical simulations exhibits a noticeable variability of 
the degree distribution with $m$. In particular, the
degree exponent strongly depends on $m$ for the general case
considered in Ref.~\cite{klemm02}. The model topology is also
susceptible to several details of the construction algorithm. By means
of large-scale numerical simulations we study the deactivation model
topology in the whole range of $m$ and for different algorithm
parameters.  We calculate analytically the clustering coefficient and
connectivity correlation functions. Also in this case a 
variability with respect to the model parameters is found. Extensive
numerical simulations confirm the analytical picture presented here.

In the generated networks, we also report the lack of small-world
properties.  In the whole parameters range we find a network diameter
increasing linearly with the number of nodes forming the network
\cite{klemm02b}.  The networks' topology is therefore similar to an
array of connected star-shaped graphs. The networks are thus similar
to a one dimensional lattice in what concerns their physical
properties.  In particular, diffusion and spreading processes might be
heavily affected by the increasing average distance among nodes that
make the system alike to a one-dimensional chain.  In this
perspective, we discuss the properties of epidemic spreading and
resilience to damage in networks generated with the deactivation
model.

\section{Deactivation model}

The deactivation model introduced by Klemm and Egu{\'\i}luz \cite{klemm02}
is defined as follows: Consider a network with directed links. Each
node can be in two states, either active or inactive. The model starts
from a completely connected graph of $m$ active nodes and proceeds by
adding new nodes one by one. Each time a node is added:
\begin{enumerate}
\item It is connected to all active nodes in the network;
\item One of the active nodes is selected and set inactive with
  probability
\begin{equation}
  p_d(\kin_i)=\frac{\left[\sum_{j\in{\cal A}}(a+\kin_j)^{-1}\right]^{-1}}{a+\kin_i};
  \label{eq:1}
\end{equation}
\item The new node is set active.
\end{enumerate}
The sum in Eq.~(\ref{eq:1}) runs over the set of active nodes ${\cal
  A}$, $a$ is a model parameter, and $\kin_i$ denotes the in-degree
of the $i$-th node.

As we shall show below, this model is quite sensitive to the order in
which steps 2 and 3 are performed and, therefore, it is better to
discriminate the following cases.
\begin{itemize}
\item {\em Model A}: step 2 is performed \textit{before} step 3.
\item {\em Model B}: step 2 is performed \textit{after} step 3.
\end{itemize}
For $m\to\infty$ both models can be solved analytically in the continuous
$\kin$ approximation, after introducing the probability density that an
active node has in-degree $\kin$~\cite{klemm02}. Moreover in this
limit the order of steps 2 and 3 is irrelevant, obtaining the same
in-degree distribution $P(\kin)\sim (a+ \kin)^{-\gamma}$ with
\begin{equation}
\gamma=2+\frac{a}{m}.
\label{eq:1a}
\end{equation}
The model is usually simulated by using $a=m$. In this way the
deactivation probability is inversely proportional to the total 
connectivity of the nodes $(m+ \kin)^{-1}$ and the 
connectivity distribution results
to be $P(k)=2m^2 k^{-3}$.  Interestingly, due to the deactivation
mechanism the networks show a high clustering coefficient that
approaches a constant value in the infinite size limit~\cite{klemm02}.
 
At lower values of $m$, it has been claimed that finite size effects
set in and the connectivity distribution shows deviations from the
predicted behavior.  We shall see in the next section that for $a=m\leq
10$ the model presents a very different analytical solution which
yields a connectivity distribution very far from the $m\to\infty$ limit. In
addition, the deactivation model topology is very sensible to changes
in the details of the growing algorithms.

\section{Degree distribution}

\subsection{Model A}
 
Let us first focus on model A with $m=2$, that is the smallest value
of $m$ for which the model is non-trivial.  In this case, after adding
a new node we have only two nodes at the deactivation step. One of
them will be set inactive and replaced by the new added node, that has
in-degree $0$.  In the worst case, the other node will have in-degree
$0+1$, the $0$ coming from its initial in-degree and the $1$ from the
connection to the newly added node, and in general it will have
in-degree larger than or equal to $2$.  Later on, at the next
deactivation step, the in-degrees of both nodes will have increased by
one resulting in one active node with in-degree $1$ and another with
in-degree $K\geq2$, where $K$ is the in-degree of the active node with
largest in-degree, that coincides with the oldest node.  Then,
following Eq.~(\ref{eq:1}), one of them will be deactivated with
probability
\begin{equation}
  p_d(K)=\frac{1+a}{1+2a+K},\ \ \ p_d(1)=1-p_d(K).
\label{eq:2}
\end{equation}
Each time the oldest node is not deactivated its in-degree increases
by one and, therefore, the probability that the oldest node has
in-degree $K$ is just the probability that it is not deactivated in
$K-2$ steps, with running in-degree $2,\ 3, \ldots K-1$. Thus, the
probability $\tilde{P}(K)$ of creating a deactivated node of in-degree
$K$ is equal to the probability that the largest node is not
deactivated in $K-2$ steps and is deactivated in the last step, {\em
  i.e.}:
\begin{eqnarray}
\tilde{P}(K) & = &\prod_{\ell=2}^{K-1}\left[1-p_d(\ell)\right]p_d(K)
\nonumber\\
&=&\frac{\Gamma(3+2a)}{\Gamma(1+a)}\frac{\Gamma(a+K)}{\Gamma(2+2a+K)},
\label{eq:3}
\end{eqnarray}
where $\Gamma(x)$ is the standard Gamma function \cite{abramovitz}. On the
other hand, every time that the oldest node is not deactivated, the
other, with in-degree $1$, is deactivated. Hence, in the $K-1$
deactivation steps leading to the generation of a node with in-degree
$K$, $K-2$ nodes with in-degree $1$ are created. The average number of
nodes with in-degree $1$ created in the process is then
\begin{equation}
  \tilde{P}_1 = \sum_{K=2}^\infty (K-2) \tilde{P}(K) = \frac{2+a}{a}.
\end{equation}
Therefore, the in-degree distribution will be given by
\begin{equation}
 P(\kin)=C^{-1} \times \left\{ \begin{array}{cl}
                      \tilde{P}_1 & \quad\kin=1 \\ & \\
                      \tilde{P}(\kin) & \quad \kin>1
                      \end{array}\right.,
\label{eq:4}
\end{equation}
where $C$ is a constant, obtained from the normalization condition
$\sum_{\kin }P(\kin)=1$, that has the value
\begin{equation}
  C= \tilde{P}_1 + \sum_{K=2}^\infty  \tilde{P}(K) = \frac{2a+2}{a}. 
\end{equation}
From this equation, we obtain the analytic expression for the
in-degree distribution
\begin{equation}
  P(\kin)= \left\{ \begin{array}{cl}
                   \displaystyle\frac{2+a}{2+2a} & \quad\kin=1 \\ & \\
                   \displaystyle\frac{\Gamma(2+2a)}{\Gamma(a)}
                   \displaystyle\frac{\Gamma(a+\kin)}{\Gamma(2+2a+\kin)}  & 
                   \quad \kin>1 
                  \end{array}\right. .
\end{equation}
For large $\kin$ we can expand the previous expression using
Stirling's approximation, to obtain that the
in-degree distribution follows the asymptotic behavior
\begin{equation}
P(\kin)\sim {\kin}^{-\gamma},\ \ \ \ \gamma=2+a.
\label{eq:4aa}
\end{equation} 
Moreover, since the out-degree of all nodes is $m$ then the degree $k$
of a node (in-degree plus out-degree) is $m+\kin$ and will follow
the same distribution shifted by $m$. For the particular case $a=m=2$,
the degree distribution takes the form
\begin{equation}
  P(k)= \left\{ \begin{array}{cl}
                   \displaystyle\frac{2}{3} & \quad k=3 \\ & \\
                   \displaystyle\frac{120}{k(k+1)(k+2)(k+3)} &
                   \quad k>3
                  \end{array}\right.
\label{eq:14}
\end{equation}

\begin{figure}[t] 
\centerline{\psfig{file=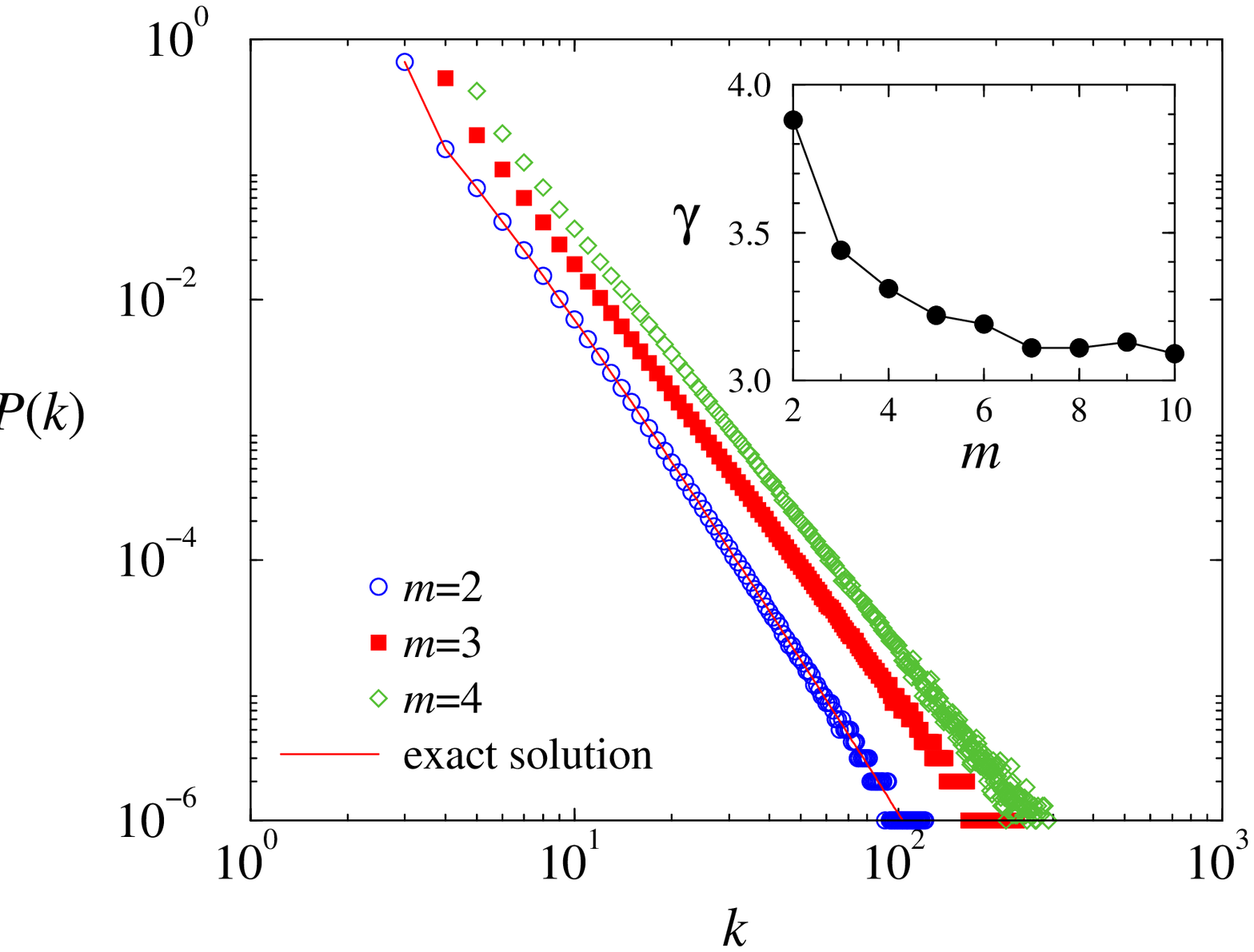,width=3in}}

\caption{Degree distribution of model A for $a=m$, network size of
  $N=10^7$, and different values of $m$. The continuous line is the
  exact distribution for $m=2$ given by Eq.~(\protect\ref{eq:14}). The
  inset shows the value of the exponent $\gamma$ as a function of $m$
  obtained from numerical simulations.}

\label{fig:1}
\end{figure}

In Fig.~\ref{fig:1} we plot the degree distribution obtained from
numerical simulations of model $A$ for $a=m$. For $m=2$ the numerical
points are in very good agreement with the exact distribution given in
Eq.  (\ref{eq:14}), with a power law decay with exponent $\gamma=2+a=4$.
In the limiting case of large $m$ the continuous approach predicts the
exponent 3~\cite{klemm02} (see Eq. (\ref{eq:1a})), giving us a lower
bound.  Hence,
\begin{equation}
\text{model A with $a=m$} \quad \Longrightarrow \quad  3<\gamma\leq4
\label{eq:4aaa}
\end{equation}
and, therefore, the degree distribution has always a bounded second moment.
For larger $m$ the distribution follows a power law decay but
with an exponent $\gamma$ that depends on $m$.
In order to show that the degree distribution approaches for each $m$ an 
asymptotic power law behavior with $\gamma> 3$ we performed 
large scale simulations  of networks with $N=10^7$ nodes.
In Fig.~\ref{fig:1} we report the behavior of the exponent $\gamma$ 
as a function of $m$. For all values of $m<10$ the degree exponents 
strongly deviates from the $m\to\infty$ limit.

\subsection{Model B}
\label{sec:model-b}
Using similar arguments we can compute the degree distribution of
model B for $m=1$. In this case we also have two nodes at the
deactivation process, the one just added and the one surviving from
the previous deactivation step. The former has in-degree $0$ while the
latter (the oldest) has in-degree $K\geq1$, and one of them is
deactivated with probability
\begin{equation}
p_d(K)=\frac{a}{2a+K},\ \ \ p_d(0)=1-p_d(K).
\label{eq:4a}
\end{equation}
The probability that when the oldest node is deactivated it has
degree $K$ is given by
\begin{eqnarray}
\tilde{P}(K) & = &\prod_{\ell=1}^{K-1}\left[1-p_d(\ell )\right]p_d(K)
\nonumber\\
&=&\frac{\Gamma(1+2a)}{\Gamma(a)}\frac{\Gamma(a+K)}{\Gamma(1+2a+K)}.
\label{eq:5}
\end{eqnarray}
In the process of creating a node of in-degree $K$, $K-1$ nodes of
in-degree $0$ have been created. The average number of nodes
with in-degree $0$ created is
\begin{equation}
  \tilde{P}_0 = \sum_{K=1}^\infty (K-1) \tilde{P}(K) = \frac{a+1}{a-1}.
\end{equation}
Thus, the analytic expression for the normalized in-degree
distribution is given by
\begin{equation}
P(\kin)=C^{-1} \times \left\{ \begin{array}{cl}
                      \tilde{P}_0 & \quad\kin=0 \\ & \\
                      \tilde{P}(\kin) & \quad \kin>0
                      \end{array}\right.,
\end{equation}
with the normalization constant
\begin{equation}
  C= \tilde{P}_0 + \sum_{k=1}^\infty  \tilde{P}(K) = \frac{2a}{a-1}.
\end{equation}
From here follows the expression for the degree distribution (where
$k=m+\kin$)
\begin{equation}
  P(k)= \left\{ \begin{array}{cl}
                   \displaystyle\frac{1+a}{2a} & \quad k=1 \\ & \\
                   \displaystyle\frac{\Gamma(2a)}{\Gamma(a-1)}
                   \displaystyle\frac{\Gamma(a+k-1)}{\Gamma(2a+k)}  & 
                   \quad k >1
                  \end{array}\right. .
                \label{eq:6}
\end{equation}
For large $k$ the degree distribution follows the asymptotic
behavior
\begin{equation}
P(k)\sim k^{-\gamma},\ \ \ \ \gamma=1+a.
\label{eq:6a}
\end{equation}

\begin{figure}[t]
\centerline{\psfig{file=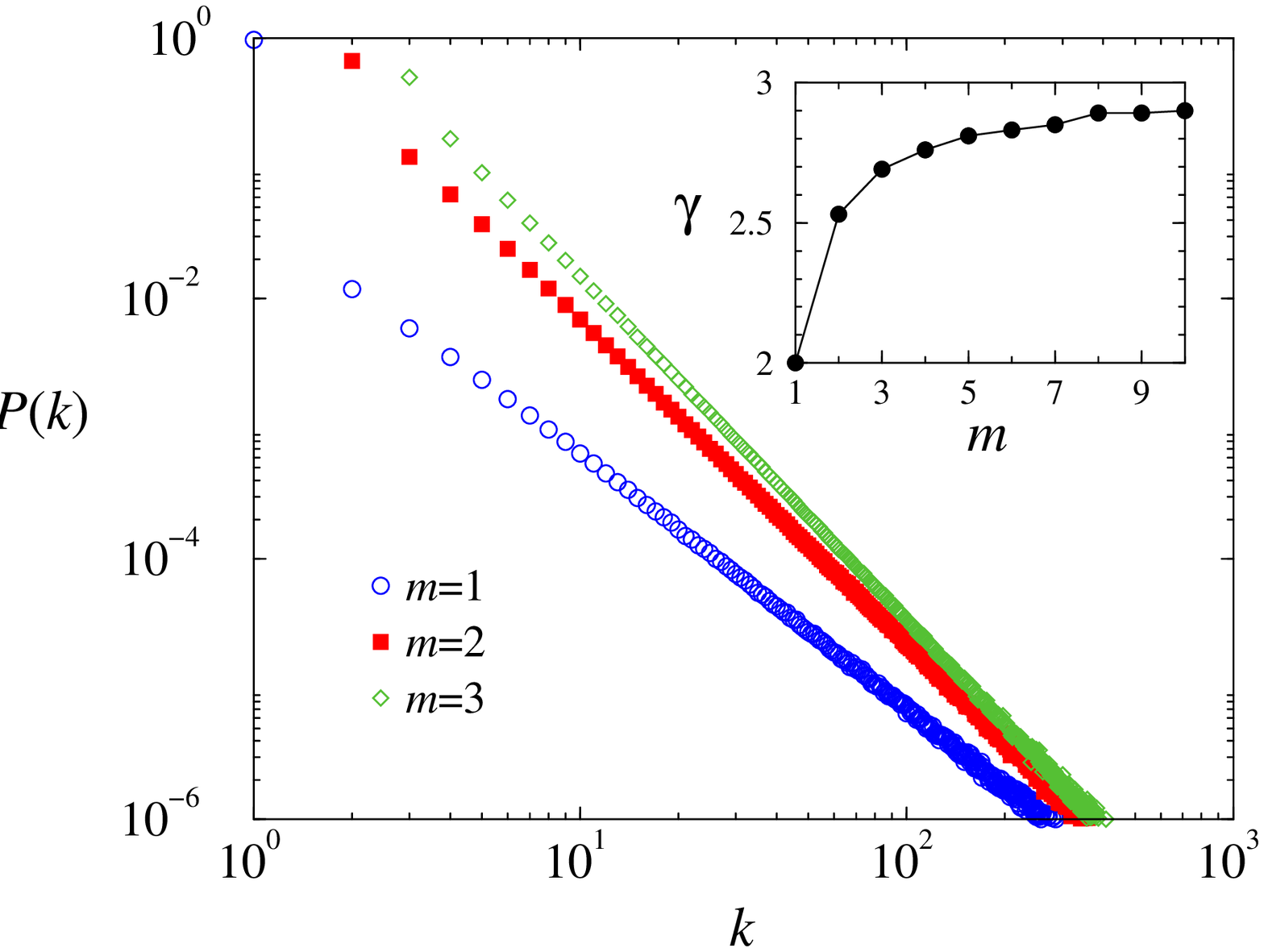,width=3in}}
\caption{Degree distribution of model B for $a=m$, network size of
  $N=10^7$ and different values of $m$. The inset shows the value of
  the exponent $\gamma$ as a function of $m$ obtained from numerical
  simulations.}
\label{fig:2}
\end{figure}

In Fig.~\ref{fig:2} we show the degree distribution obtained from
numerical simulations of model B with $a=m$.  For $a=m=1$ we recover
the predicted exponent $\gamma=-2$.
Also in this case, we provide large scale numerical simulations ($N=10^7$) 
of networks with larger values of $m$. The obtained distributions 
still follow a power law decay but with an exponent $\gamma$ that is a
continuously  increasing function of $m$. It is worth remarking that
for $m < 10$ the degree  exponent is stable and strongly differs from 
the value $\gamma=3$. 

\begin{figure}[t]
\centerline{\psfig{file=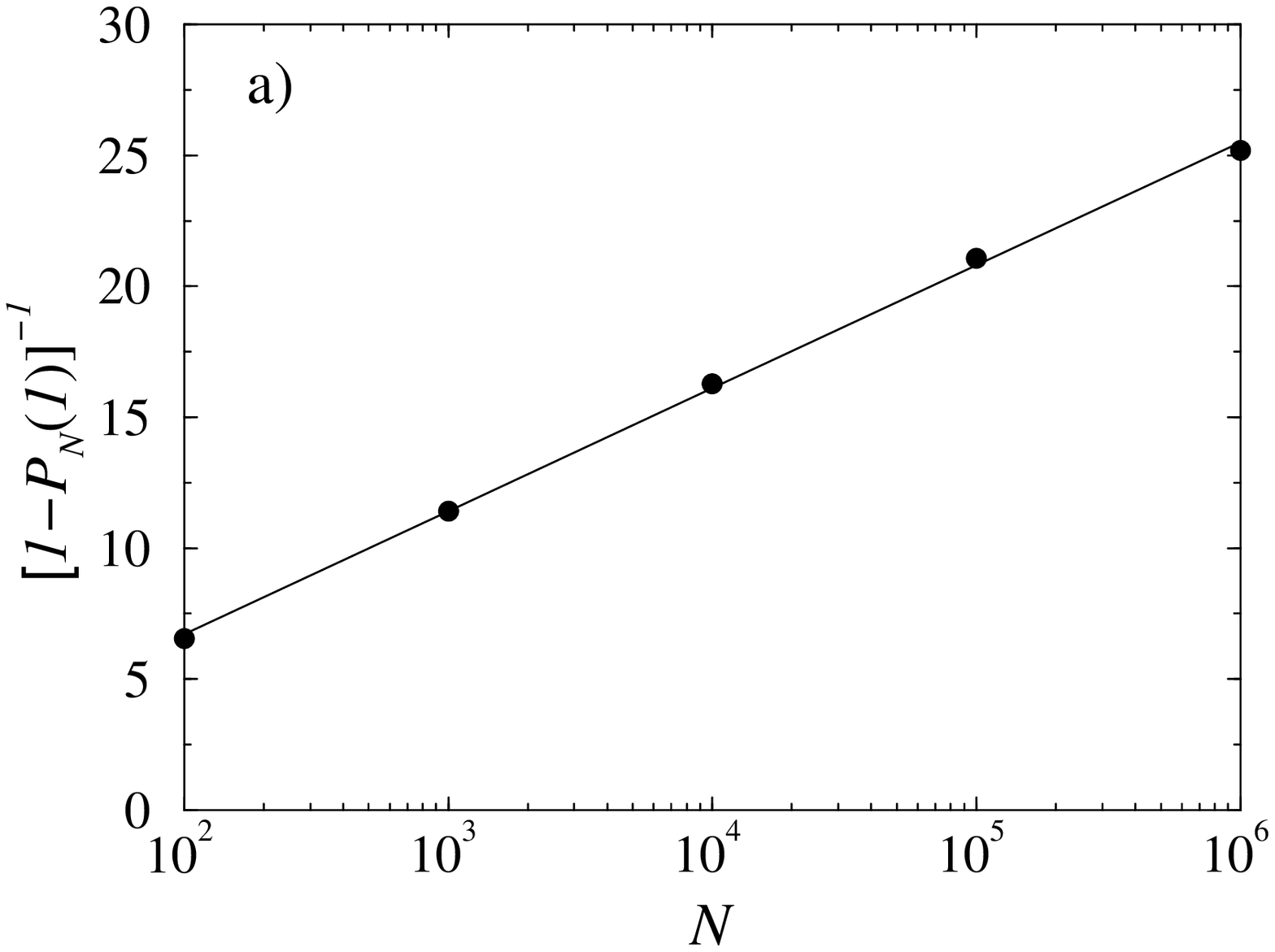,width=3in}}
\centerline{\psfig{file=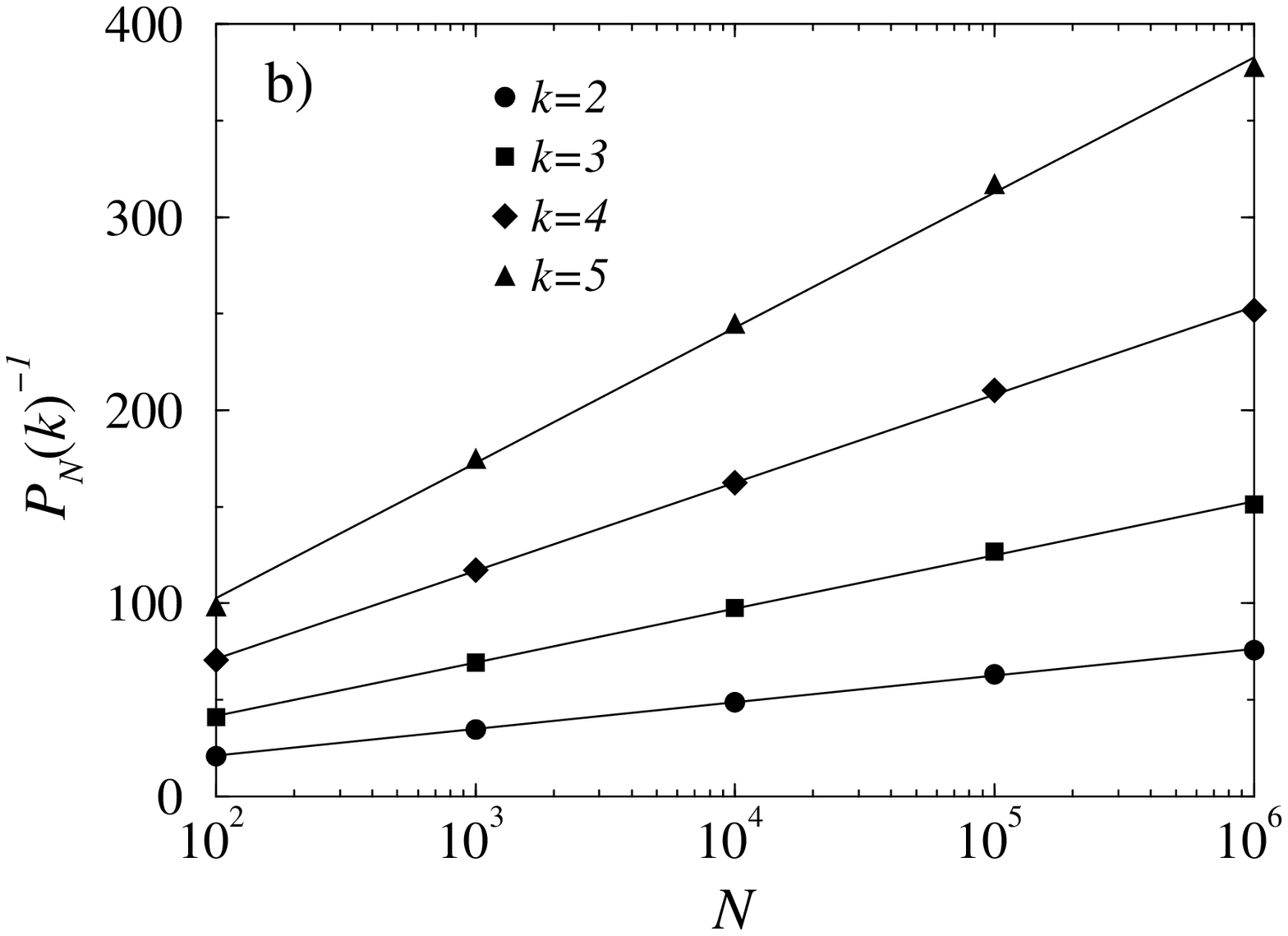,width=3in}}
\caption{Scaling of the degree distribution $P_N(k)$ for the B model
  with $a=m=1$ at fixed $k$, as a function of the network size $N$,
  for (a) $k=1$ and (b) $k>1$.  The solid lines are least-squares fits
  to the form $[1-P_N(1)]^{-1} \sim \ln N$ in (a) and  $P_N(k)^{-1} \sim \ln
  N$ in (b), as predicted by Eq.~(\protect\ref{eq:17})}
\label{fig:scaling}
\end{figure}

It is worth noticing that for $a=m=1$ the analytic solution
Eq.~(\ref{eq:6}) is singular, as can be readily seen from the
$\Gamma(a-1)$ factor in the denominator. In fact, the solution in this
case is $P(k) =\delta_{k,1}$, that is, in the thermodynamic limit
(infinitely large network), the weight of the nodes with degree $1$ is
overwhelming with respect to the nodes with different connectivity.
This singularity is rooted in the fact that the distribution, with
exponent $\gamma=-2$, lacks a finite first moment in the thermodynamic
limit, while we know that, by definition, model $B$ has average
connectivity $\avk=2$. This necessarily implies that there must be an
implicit dependence on the network size $N$ in the degree distribution
for $a=m=1$, dependence that cannot be assessed by our analytic
solution since we are already working in the infinite network limit.
We can nevertheless estimate the functional form of the degree
distribution for a finite network composed by $N$ nodes, which has a
maximum connectivity $k_c$, such that there are no nodes with degree
larger that $k_c$. Assuming that the distribution for $k>1$ follows
the same functional form as Eq.~(\ref{eq:6}), we have that for $a=1$
\begin{equation}
  P_N(k)= \left\{ \begin{array}{cl}
                   C_1 & \quad k=1 \\ & \\
                   \displaystyle\frac{C_2}{k(k+1)} &
                   \quad 1<k\leq k_c
                  \end{array}\right. ,
\label{eq:15}
\end{equation}
where $C_1$ and $C_2$ are constants to be determined by the
normalization conditions $\sum_{k=1}^{\infty} P_N(k) = 1$ and $\sum_{k=1}^{k_c}
k P_N(k) = 2$ (the upper limit in the first normalization condition can
be taken to be infinite, since the corrections stemming from $k_c$ are
of lower order). From this two conditions we obtain, in the continuous
$k$ approximation that replaces sums by integrals,
\begin{equation}
  C_1 = 1- \frac{2
    \ln(3/2)}{\ln\left(\displaystyle\frac{1+k_c}{2}\right)}, \qquad  
  C_2 = \frac{2}{\ln\left(\displaystyle\frac{1+k_c}{2}\right)}.
\end{equation}
For finite SF networks with degree distribution $P(k) \sim k^{-\gamma}$, the
maximum degree $k_c$ scales with the number of nodes as $k_c \sim
N^{1/(\gamma-1)}$ \cite{dorogorev}. In the present case we have $k_c \sim
N$, and thus, for large $N$,
\begin{equation}
  1- C_1 \sim  \frac{1}{\ln N}, \qquad   C_2 \sim  \frac{1}{\ln N}.
\label{eq:16}
\end{equation}
Therefore, in the limit $N\to\infty$, we recover a singular degree
distribution with $C_1 \to 1$ and $C_2 \to 0$. We can check numerically
this result by noticing that, from Eqs.~(\ref{eq:15})
and~(\ref{eq:16}), the degree distribution at fixed $k$ should scale
as
\begin{equation}
  1-  P_N(1) \sim  \frac{1}{\ln N}, \quad P_N(k) \sim \frac{1}{\ln N} \;
  ~~k>1. 
\label{eq:17}
\end{equation}
We have verified this scaling form in Fig.~\ref{fig:scaling}.
Therefore, in model B with $a=m=1$ we obtain a degree distribution
that decays as $k^{-2}$, but with a normalization constant for $k>1$
that decays with the network size as $1/ \ln N$.  
Finally, it is worth mentioning that the second moment of the 
distribution is diverging as $\fluck\sim N/\ln N$.
Despite this singular
behavior for $a=m=1$, however, Eq.~(\ref{eq:6}) remains exact for any
value of $a \neq 1$.

From the results of Fig.~\ref{fig:2}, together with the upper bound
$\gamma=3$ obtained from the large $m$ approximation~\cite{klemm02}, we
have that
\begin{equation}
\text{model B with $a=m$}  \quad \Longrightarrow \quad  2\leq\gamma<3,
\label{eq:6aa}
\end{equation}
and, therefore, the degree distribution has a divergent second moment.

The analysis made above has shown that the deactivation model is quite
sensitive to the order in which steps 2 and 3 are performed, yielding
degree distributions with a finite or divergent second moment,
depending on the order. In addition, the exponent $\gamma$ is rather
sensible to the value of $a=m$, showing a wide range of variation.
This fact has not been noticed in previous works where this model has
been considered \cite{klemm02,structured,crucitti02}, prompting that
some of the conclusions obtained in those works should be reconsidered
in this perspective.

\section{Clustering coefficient}

We can go beyond the degree distribution and compute the clustering
coefficient $c(k)$, as a function of the node degree $k$
\cite{alexei02,klemm02b}. For this quantity we can perform an analytic
calculation for any value of $a$ and $m$ and for both models A and B.
In order to compute the clustering coefficient, we will consider the
network as undirected and denote by $k_i=\kin_i+m$ the total degree of
the node $i$.

The clustering coefficient of the node $i$ is defined by
\cite{watts98}
\begin{equation} 
c_i=\frac{2e_i}{k_i(k_i-1)}, 
\label{eq:7} 
\end{equation}
where $e_i$ is the number of edges between the neighbors of node $i$
and it is divided by its maximum possible value $k_i(k_i-1)/2$. In the
deactivation model new edges are created between the active nodes and
the added node. Hence, $e_i$ remains constant for inactive nodes and
increases only for the active ones. Moreover, all the active nodes are
connected. Hence, each time we add a node the degree $k_i$ of each
active node $i$ increases by one and $e_i$ increases by $m-1$, where
$m-1$ are just the new links between the new neighbor of $i$ (the
added node) and the remaining active nodes. Therefore, the dynamics of
$e_i$ is given by
\begin{equation} 
\frac{\partial e_i}{\partial t}=(m-1),
\label{eq:8}
\end{equation}
while the connectivity obeys the relation $k_i(t)=m+t$. Here $t=0$
corresponds to the time at which the node $i$ was created.  Besides,
when the node is added it has degree $m$, thus $e_i(0)=m(m-1)/2$ and,
therefore, $c_i(0)=1$. Integrating Eq. (\ref{eq:8}) with this initial
condition and substituting the result in Eq. (\ref{eq:7}), taking into
account that $t=k_i-m$, we obtain
\begin{eqnarray}
  c(k)&=& \frac{m(m-1)}{k(k-1)}+\frac{2(m-1)(k-m)}{k(k-1)}
  \nonumber \\
  &=&\frac{2(m-1)}{k}-\frac{(m-1)(m-2)}{k(k-1)},
\label{eq:9}
\end{eqnarray}
where the last expression in Eq.~(\ref{eq:9}) is obtained after some
algebraic manipulations. Eq.~(\ref{eq:9}) recover the result
previously obtained in Ref.\cite{klemm02b}. For $m=1$ the network is a
tree, and therefore we obviously recover $c(k)=0$. For $m=2$ we obtain
the exact behavior $c(k)= 2/k$. For $m>2$, the asymptotic behavior for
large $k$ is $c(k)\sim 1/k$ \cite{klemm02}. Interestingly, we recover in
this model the same behavior of $c(k)$ found in other systems in
Ref.~\cite{ravasz02}.

\begin{figure}[t]

\centerline{\psfig{file=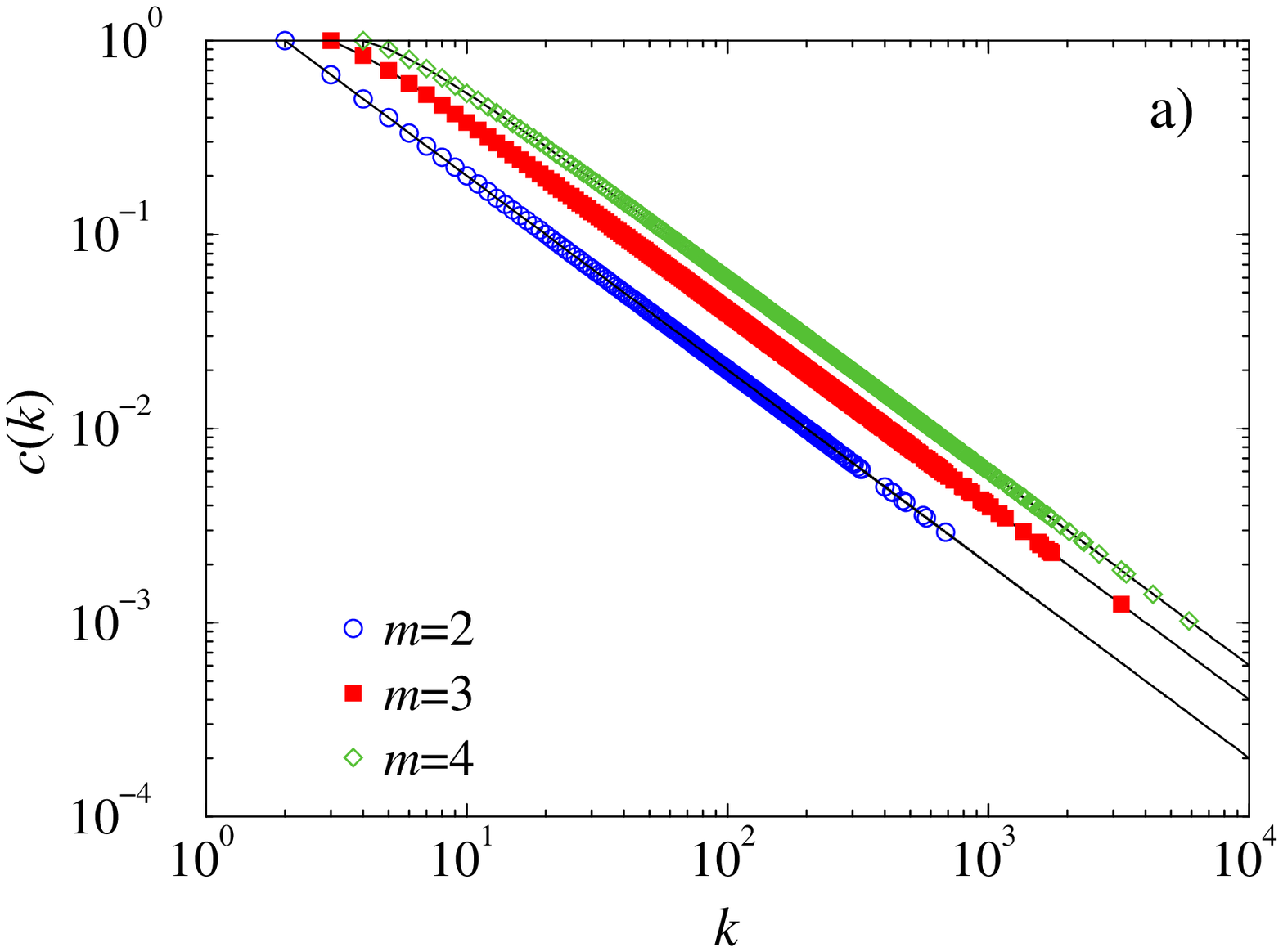,width=3in}}

\centerline{\psfig{file=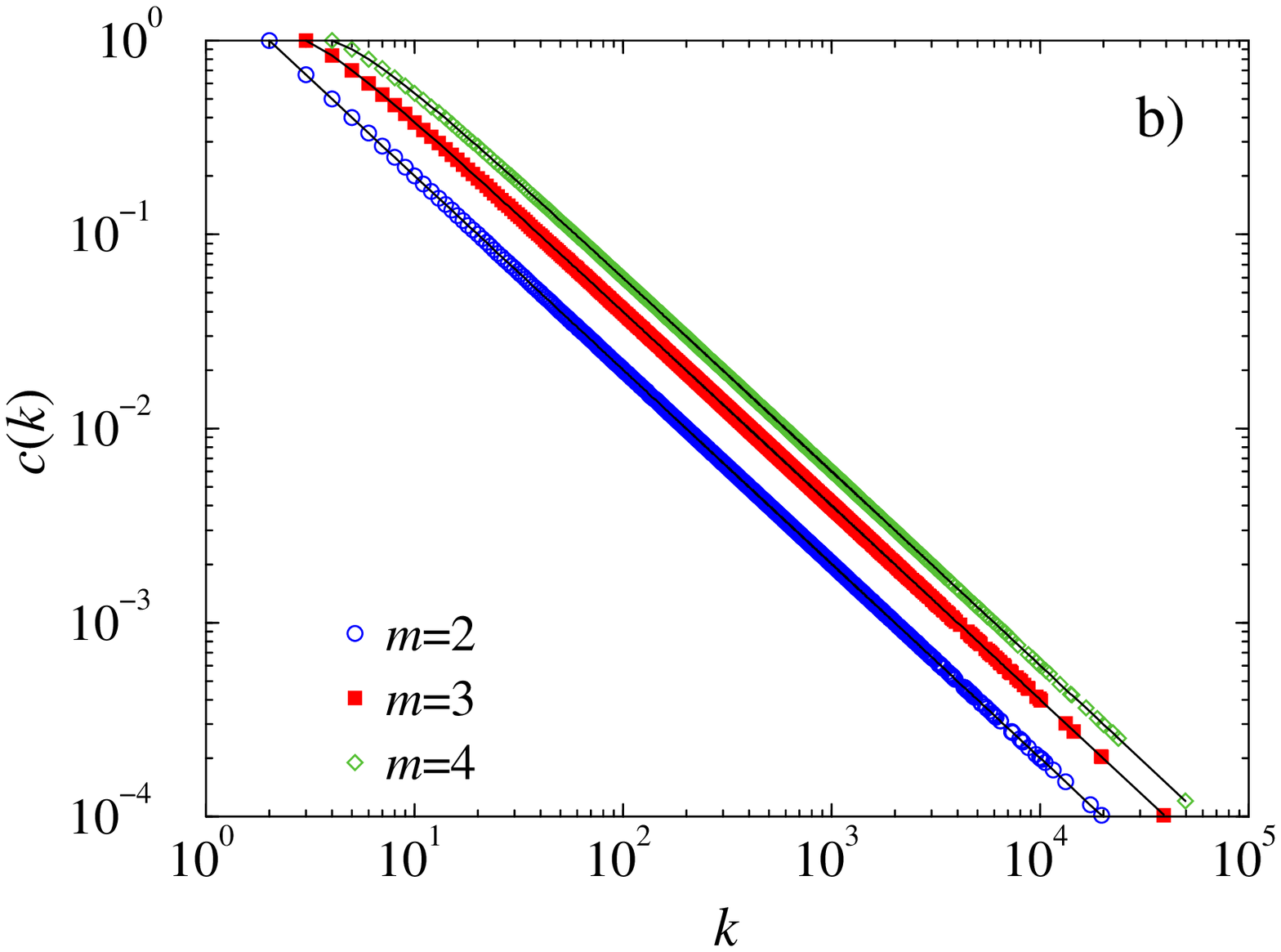,width=3in}}

\caption{Clustering coefficient as a function of the node degree for
different values of $m$. The points were obtained from numerical
simulations of (a) model A and (b) model B, up to a network size
$N=10^5$. The continuous lines correspond with the analytical solution
given in Eq.~(\protect\ref{eq:9})}

\label{fig:3}
\end{figure}

In Fig.~\ref{fig:3} we plot the clustering coefficient as a function
of the node degree obtained for model A and B and different values of
$m$ from numerical simulations. As it can be seen the numerical
dependency coincides with the analytical expression in
Eq.~(\ref{eq:9}).

\section{Degree correlation function}

Degree correlations can be characterized by analyzing the nearest
neighbor average degree introduced in Refs.~\cite{alexei,alexei02},
defined as
\begin{equation}
  k_{nn,i}=\frac{D_i}{k_i},
\label{eq:10}
\end{equation}
where $D_i$ is the sum of the degrees of the neighbors of node $i$.
In uncorrelated networks, the quantity $k_{nn,i}$ does not show any
dependence on the degree of the node $i$. This is not the case when
degree correlations are present. In this case $k_{nn,i}$ is a function
of the degree of the node whose nearest neighbors are analyzed. In
particular, we can face two possible kinds of correlation.  In the
first situation, nodes with high connectivity will connect more
preferably to highly connected nodes; a property referred to as
``assortative mixing''. On the opposite side, it is possible to have
``dissortative mixing''; i.e. highly connected nodes are preferably
connected to nodes with low connectivity \cite{assortative}.

In the deactivation model, when the node is added it has degree $m$
and $D_i=m\left<k\right>_{\cal A}$, where $\left<k\right>_{\cal A}$ is
the average degree among active neighbors. Then, if the node $i$ is
active, it is, by construction, neighbor of the $m-1$ remaining active
nodes.  Thus, every time a new node is added, $k_i$ increases by one
and $D_i$ increases by $(m-1)+m$, the $m-1$ because the degree of the
remaining $m-1$ neighbors have also increase by one and the $m$
because the new neighbor has degree $m$. Hence
\begin{equation}
\frac{\partial D_i}{\partial t}=(2m-1),
\label{eq:11}
\end{equation}
for each active node $i$. Integrating this equation, taking into
account the initial condition $D_i(0) = m\left<k\right>_{\cal A}$ and
the relation $t=k_i-m$, we obtain that 
\begin{equation}
  D_i'=(2m-1)(k_i-m)+m\left<k\right>_{\cal A}
\label{eq:11a}
\end{equation}
when the node $i$ is deactivated.  Now, when an active node becomes
inactive its degree remains fixed but the degree of its active
neighbor nodes will still increase until they get deactivated.
Therefore, in the infinite time limit, we have
\begin{equation}
  D_i = D_i' + \Delta D_i,
  \label{eq:11b}
\end{equation}
where $\Delta D_i$ is the increase of $D_i$, since node $i$ was set
inactive until all its neighbors are set inactive. 

Hence, from Eqs. (\ref{eq:10}), (\ref{eq:11a}), and (\ref{eq:11b}), it
follows that
\begin{equation}
  k_{nn,i}=2m-1+\frac{m\left<k\right>_{\cal A}+\Delta D_i-m(2m-1)}{k_i}.
\label{eq:12}
\end{equation} 
It remains now the task to assess the possible dependence of $\Delta D_i$
on the connectivity $k_i$ (it is clear that the long time average of
$\left<k\right>_{\cal A}$ must be independent of the connectivity of
any deactivated node). For the minimum $m$ ($m=2$ for model A and
$m=1$ for model B) the degree of an active node set inactive is not
correlated with the degree of the remaining active nodes, since those
remaining nodes have always degrees $2$ and $3$ in model A with $m=2$,
and degree $1$ in model B with $m=1$, independently of the degree of
the last deactivated node. Therefore in this case $\Delta D_i$ cannot
depend on $k_i$. This lack of correlations is also clear for $m \gg 1$
where the sum $\sum_{j\in{\cal A}}(a+k_j)^{-1}$ in Eq. (\ref{eq:1}) is a
constant \cite{klemm02} and, therefore, the degree of the active nodes
in not correlated with the degree of the inactive nodes.  For
intermediate values of $m$, however, the degree of the active nodes
may be correlated in such a way that $\Delta D_i$ depends on $k_i$.

\begin{figure}[t]

\centerline{\psfig{file=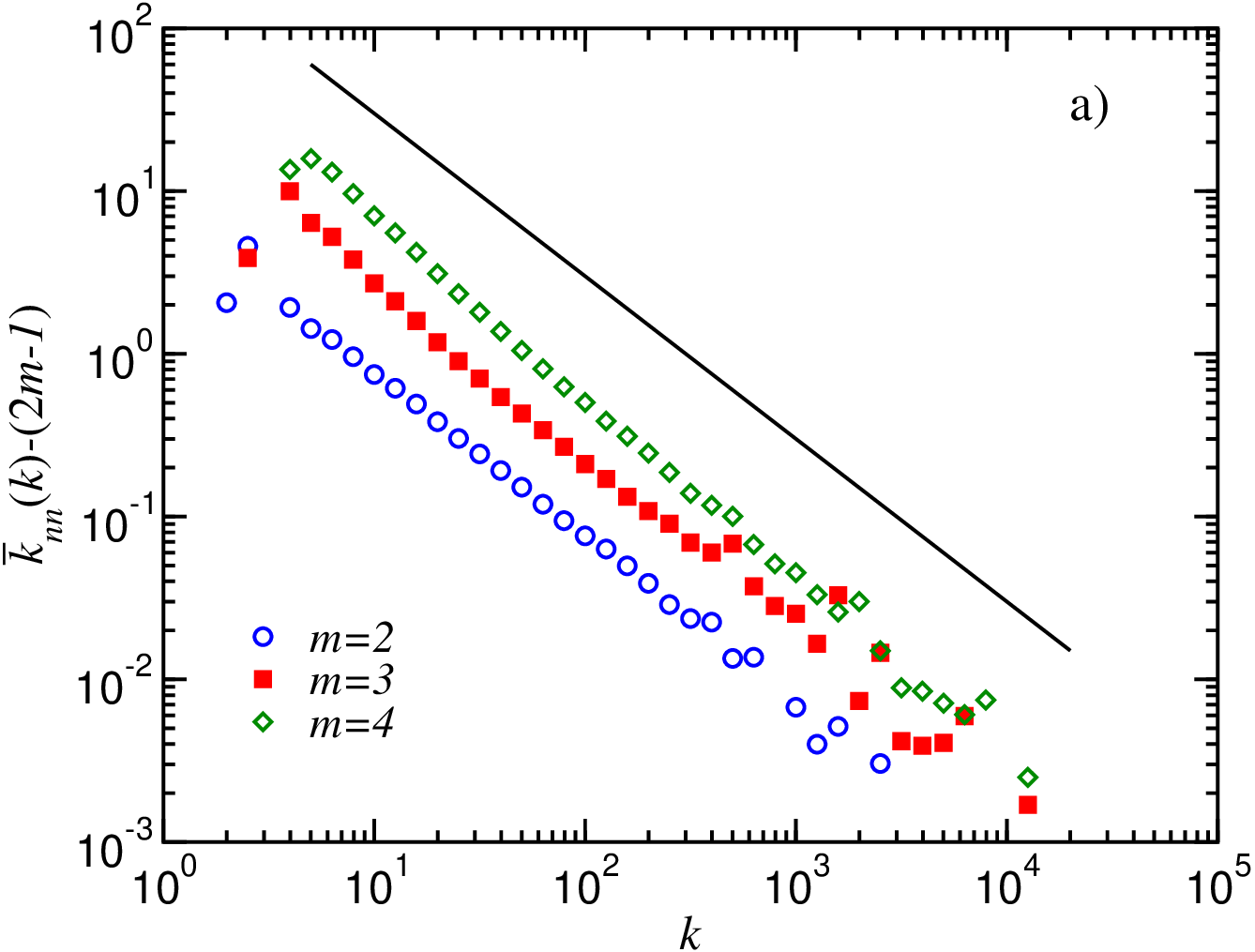,width=3in}}
\vspace*{1cm}
\centerline{\psfig{file=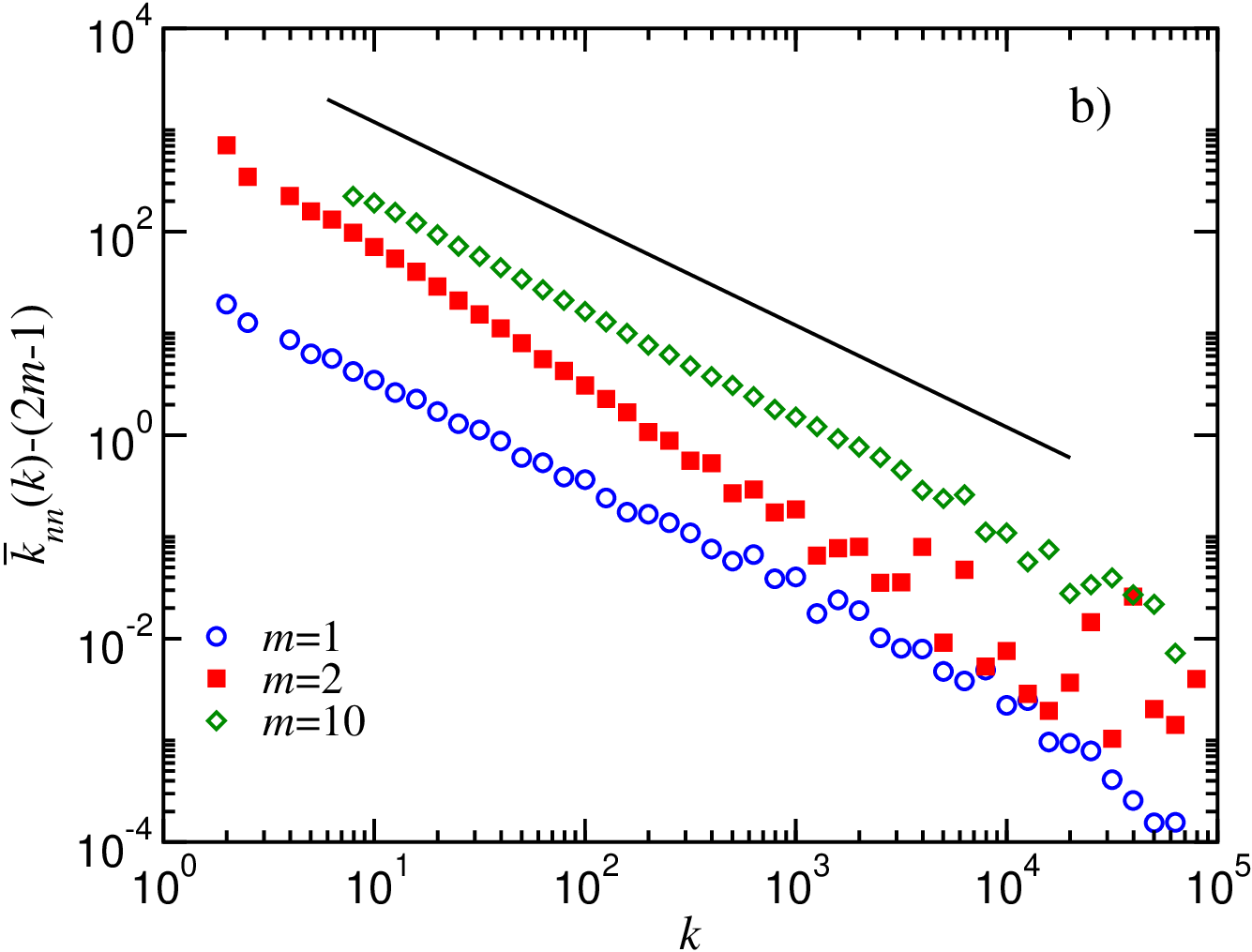,width=3in}}

\caption{Average nearest neighbor degree as a function of the degree
  $k$ for different values of $m$. The points were obtained from
  numerical simulations of (a) model A and (b) model B, up to a
  network size $N=10^5$, averaging over 1000 realizations. The
  continuous lines correspond with the analytical dependency
  $\bar{k}_{nn}(k)-(2m-1)\sim1/k$.}

\label{fig:4}
\end{figure}

In Fig.~\ref{fig:4} we plot the dependency of the average nearest
neighbors degree $\bar{k}_{nn}(k)$ as a function of the degree $k$ for
models A and B and different values of $m$. In the case of model A,
$\bar{k}_{nn}(k)-(2m-1)\sim1/k$ even for $m\neq2$, in agreement with
Eq.~(\ref{eq:12}). In the case of model B,
$\bar{k}_{nn}(k)-(2m-1)\sim1/k$ for the $m=1$ and $m=10$ but decays
faster for intermediate values of $m$. Thus, in this case the
correlations between the active node degrees introduce stronger
deviations for intermediate values of $m$.  In all cases, however, we
find that correlations in the deactivation model are of
``disassortative'' nature; \textit{i.e.} highly connected nodes are
preferably connected with poorly connected nodes.  It is also worth
stressing that the results for model B with $m=1$ must be taken with a
grain of salt, given the singular nature of the model exposed in
Sec.~\ref{sec:model-b}.

\begin{figure}[t]

\centerline{\psfig{file=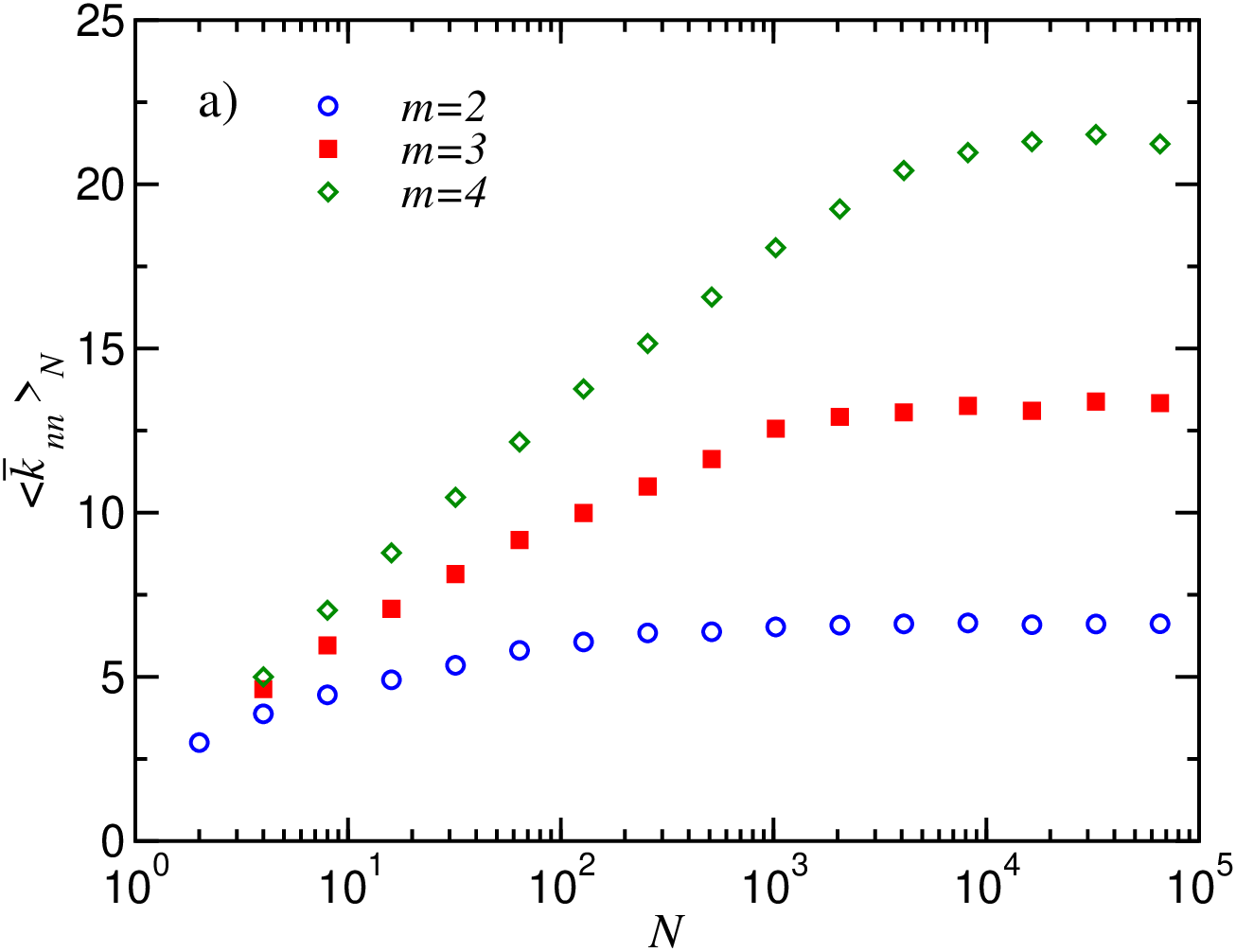,width=3in}}
\vspace*{1cm}

\centerline{\psfig{file=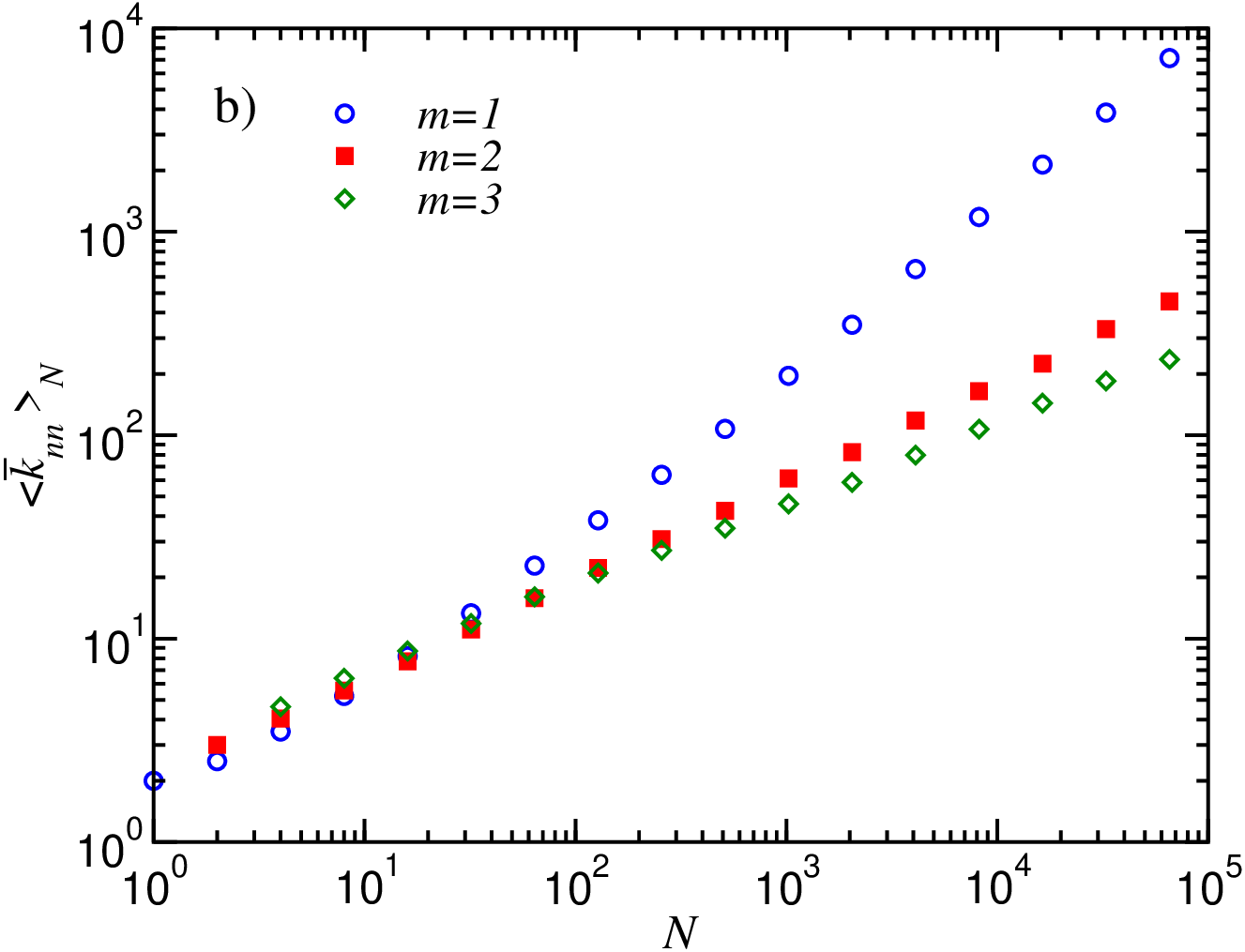,width=3in}}

\caption{Average nearest neighbor degree as a function of the network
  size $N$ for different values of $m$. The points were obtained from
  numerical simulations of (a) model A and (b) model B, up to a
  network size $N=10^5$, averaging over 1000 realizations.}

\label{fig:5}
\end{figure}

In the deactivation model, either A or B, for a fixed network size $N$
and assuming that $\Delta D_i$ does not grow faster than $k_i$, we have
that in the limit $k_i \to \infty$, $k_{nn,i} \to 2m -1$. That is, the
average nearest neighbor degree of the hubs (nodes with largest $k_i$)
equals $\avk -1$, as previously pointed out in Ref.~\cite{structured}.
However, this fact does not necessarily imply that $\Delta D_i$ is
independent of $N$. One way to check this point is to compute the
average of $k_{nn,i}$ over all nodes, $\left<\bar{k}_{nn} \right>_N =
\sum_k P(k) \bar{k}_{nn}(k)$. Let us assume that $\bar{k}_{nn} \sim \avk
-1 + \alpha / k$, where $\alpha$ is depending on $\Delta D_i$.  If $\Delta D_i$ is
approaching a constant value, we should obtain $\left<\bar{k}_{nn}
\right>_N \sim $ const., independently of $N$.  In Fig. \ref{fig:5} we
show how $\left<\bar{k}_{nn}\right>_N$ behaves with increasing $N$ for
$a=m$.  For model A, where $3<\gamma\leq4$, it approaches a stationary value
for $N\gg1$.  Moreover, the asymptotic limit of
$\left<\bar{k}_{nn}\right>_N$ increases with increasing $m$. In fact,
with increasing $m$ the exponent $\gamma$ decreases approaching the limit
$\gamma=3$ for $m\gg1$, where $\left<k^{nn}\right>$ diverges
logarithmically with $N$. On the contrary, for model B, where
$2\leq\gamma<3$, $\left<\bar{k}_{nn}\right>_N$ is growing with $N$ following
a power law. This implies that $\Delta D_i$ is a diverging function of $N$
and that in the thermodynamic limit (in which we perform first the
limit $N\to\infty$) the average nearest neighbor connectivity curve is
progressively shifting to larger and larger values. This finally
points out that the average nearest neighbor connectivity of hubs is
not a well-defined quantity since the $k_i\to\infty$ limit must be
performed only after the $N\to\infty$ limit.  The divergence of
$\left<\bar{k}_{nn}\right>_N$ with $N$ is related to a general
property of SF networks with diverging connectivity fluctuations and
it is dictated by the detailed balance of
connectivity~\cite{marian1,marian3}.

\section{Diameter and shortest path length}

Another fundamental topological feature of complex networks is
identified by the scaling of the average path length among nodes and
the network's diameter.  The minimum path between two nodes is given
by the minimum number of intermediate nodes that must be traversed to
go from node to node.  The average minimum path length
$\left<d\right>$ is thus defined as the minimum path distance averaged
over all the possible pairs of nodes in the network. Similarly, the
network diameter is defined as the largest among the shortest paths
between any two nodes in the network.

While regular networks (for instance hypercubic lattices) have a
diameter scaling with the size $N$ as the inverse of the Euclidean
dimension, many complex networks show striking small-world properties;
\textit{i.e.} in average one can go from one node to any other node in
the system by passing through a very small number of intermediate
nodes \cite{watts98}. In this case the graph diameter grows
logarithmically, or even slower, with the system's number of nodes
$N$.

\begin{figure}[t]

\centerline{\psfig{file=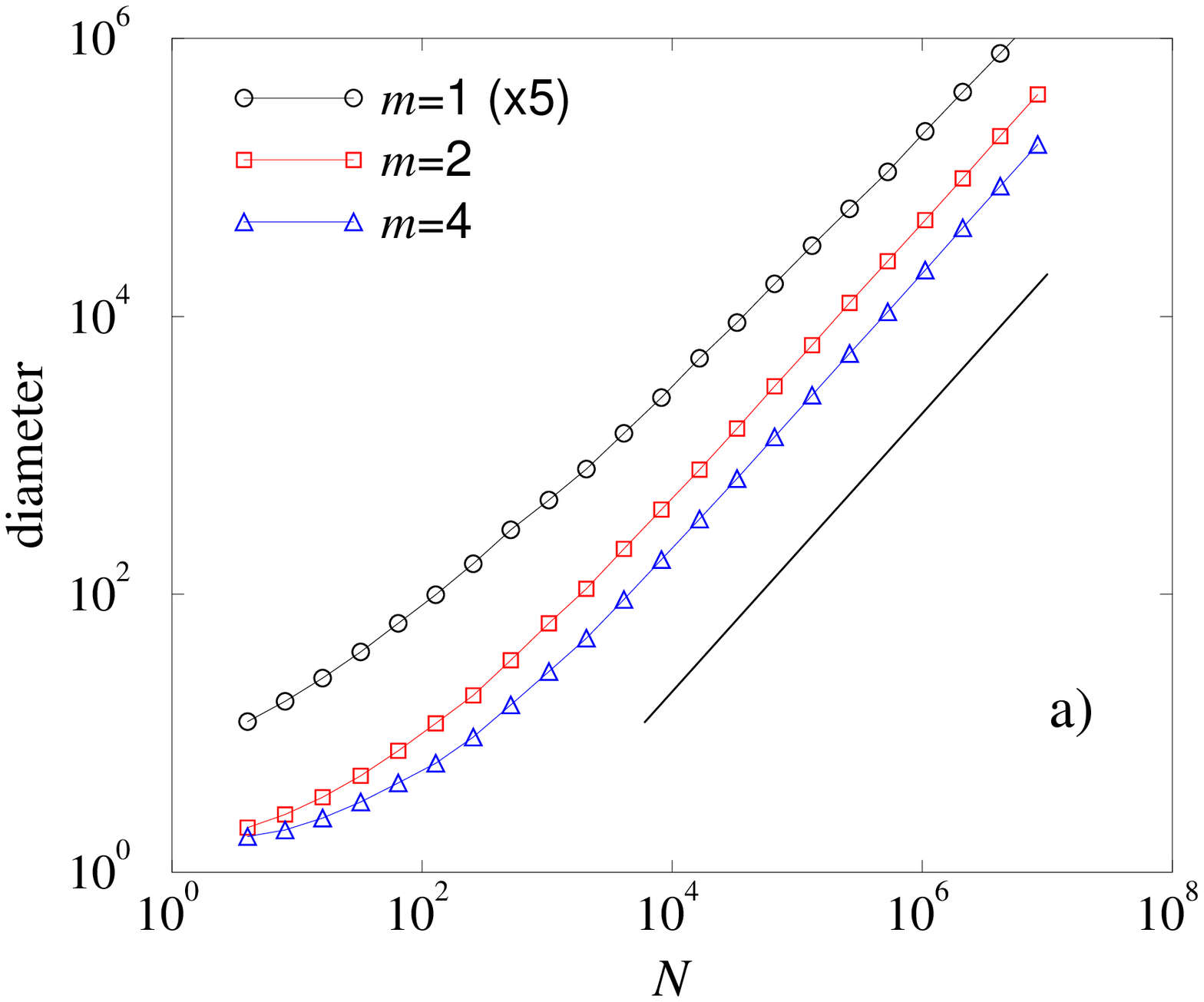,width=3in}}
\vspace*{1cm}

\centerline{\psfig{file=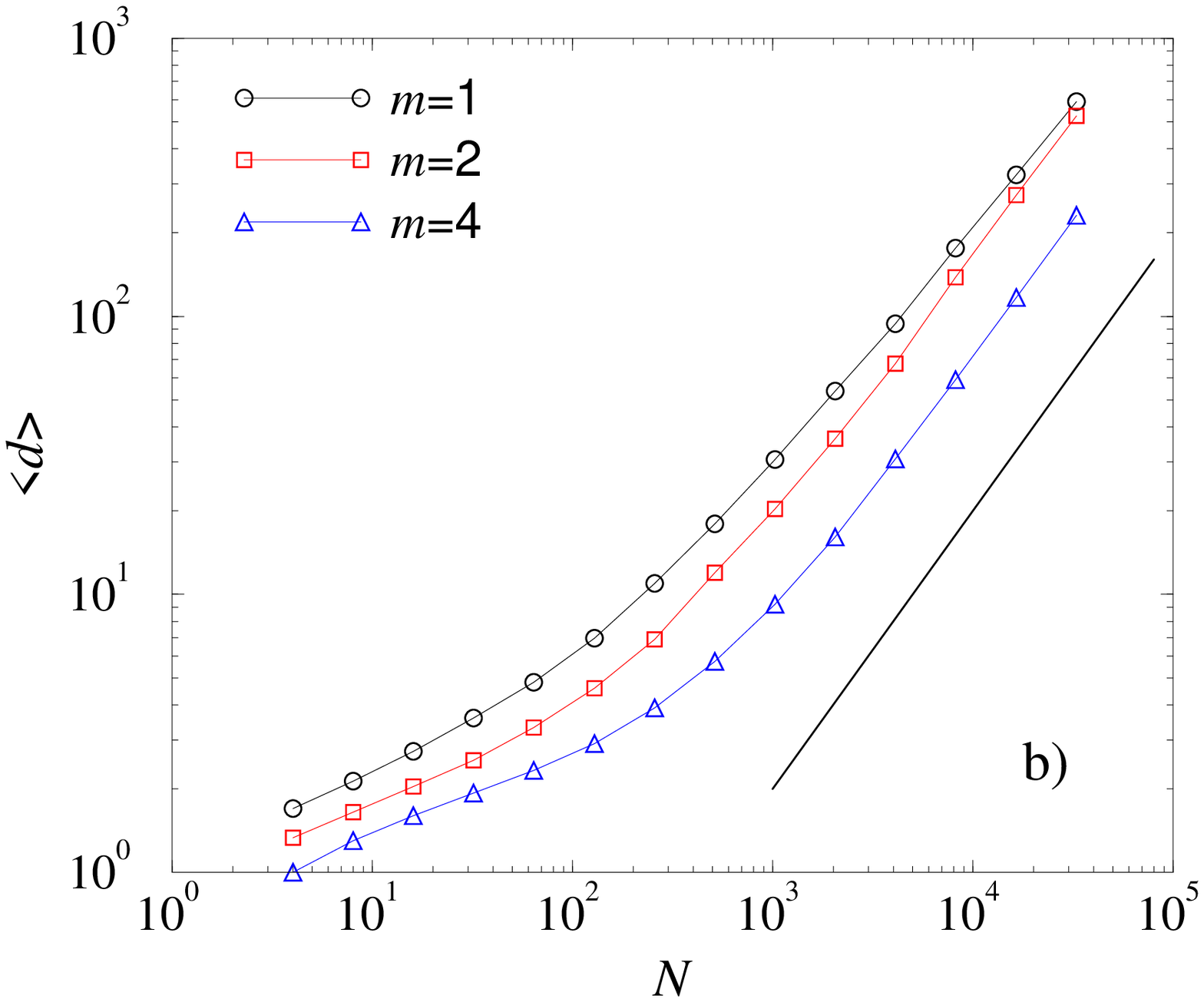,width=3in}}

\caption{Scaling of the diameter (a) and the average shortest path 
  $\left<d\right>$ (b) in the model B for different values of $m$. The
  reference lines have slope 1. For the sake of clarity, the curve
  for $m=1$ in (a) has been shifted by a factor $5$.}
\label{fig:7}
\end{figure}

In Ref.~\cite{klemm02b}, it has been noticed that for large $m$ values
$\left<d\right>$ is scaling linearly with the network size $N$.  In
the deactivation model (A and B) we measured both the diameter and the
average minimum path distance $\left<d\right>$ as a function of $N$
for values of $a=m$ ranging from $1$ to $4$. In all cases we find that
after a small size transient both metrics approach a linear scaling
with $N$. In Fig.~\ref{fig:7} we report the results obtained in the
case of the deactivation model with rule B.  This evidence implies
that the topology of the generated networks is approaching those of a
one dimensional lattice. In other words, the deactivation model does
not exhibit small-world properties.

\begin{figure}[t]

\centerline{\psfig{file=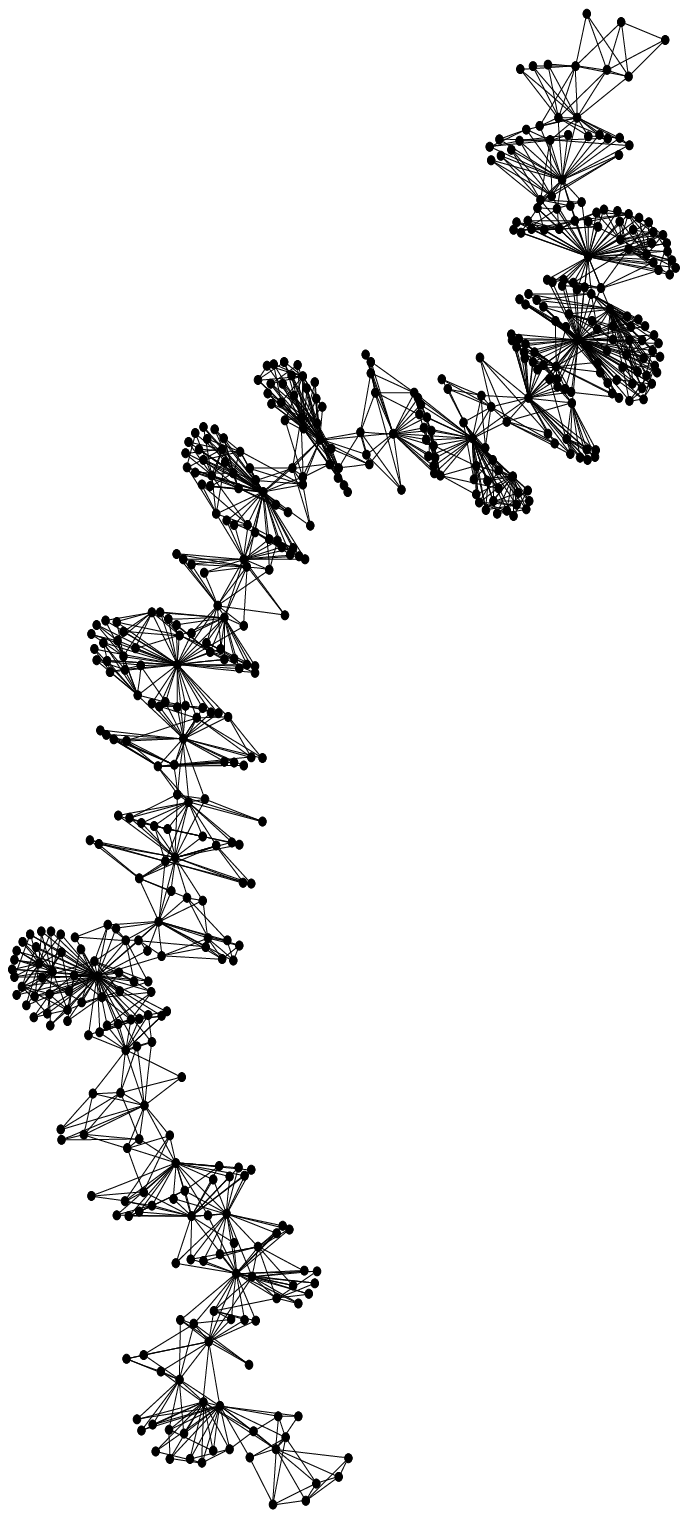,width=3in}}

\caption{Illustration of a typical network generated with the deactivation
  model B with $a=m=3$ (the size is $N=10^3$). The linear topology
  with star-like graphs connected as a chain is evident.}

\label{fig:8}
\end{figure}

In order to provide a visual representation of the deactivation model
topology, we report in Fig.~\ref{fig:8} the illustration of a network
generated with model B and $a=m=3$.  The linear structure of the
network is evident. In particular, we find a long chain of
sequentially connected star-like structures.  The star-like structures
correspond to groups of nodes connected to a node which has been
active for longer times and has had the possibility to develop a high
number of connections. Once these hubs are deactivated, they do not
receive any further connection. The network grows by adding bridge
nodes which are rapidly deactivated, until a new star-like structure
is developed by a node that is active long enough.  The growth
mechanism, however, does not allow the formation of shortcuts between
the deactivated region of the network and the new active nodes,
hindering the development of small-world properties.  The linear chain
is therefore reflecting the time evolution of the structure: recently
added nodes are separated from the original core of active nodes by a
sequence of deactivated nodes that increases proportionally to the
network size.  By inspecting networks with larger $m$ we find very
similar structures, with an increasing size of the star-like
structures forming the linear chain.  As we shall discuss in the last
section, the absence of small-world properties might have a relevant
effect in many physical properties of the network.

\section{Discussion and conclusions}

In the present work we have provided a detailed analysis of the
deactivation model introduced in Ref.~\cite{klemm02}.  The model shows
a rich behavior, being very sensible to the value of the parameters
used in the model and slight variations of the growing algorithm.  The
most striking result is that the degree distribution is depending on
the value of the number of simultaneously active nodes $m$ also in the
case in which $a=m$; \textit{ i.e.} when the deactivation
probability is related to the nodes' total degree.  The degree
exponent is asymptotically approaching the value $\gamma=3$ only for
$m\to\infty$, and the SF properties of networks suffer large variations in
the range $1\leq m\leq 10$.  Along with the high clustering observed in
previous works, we find that the model exhibits interesting degree
correlation properties. In particular, we find marked disassortative
mixing properties; \textit{i.e.} highly connected nodes link
preferably to poorly connected nodes. The analytical expression for
the degree correlation is obtained and recovered by numerical
simulations.  Strikingly, the SF and correlation properties are not
associated with small-world properties. The numerical analysis shows
that for all values of $m$ the network diameter is increasing linearly
with the number of nodes.  The network thus approaches a linear
structure, lacking long-range shortcuts.
 
\begin{figure}[t] 

\centerline{\psfig{file=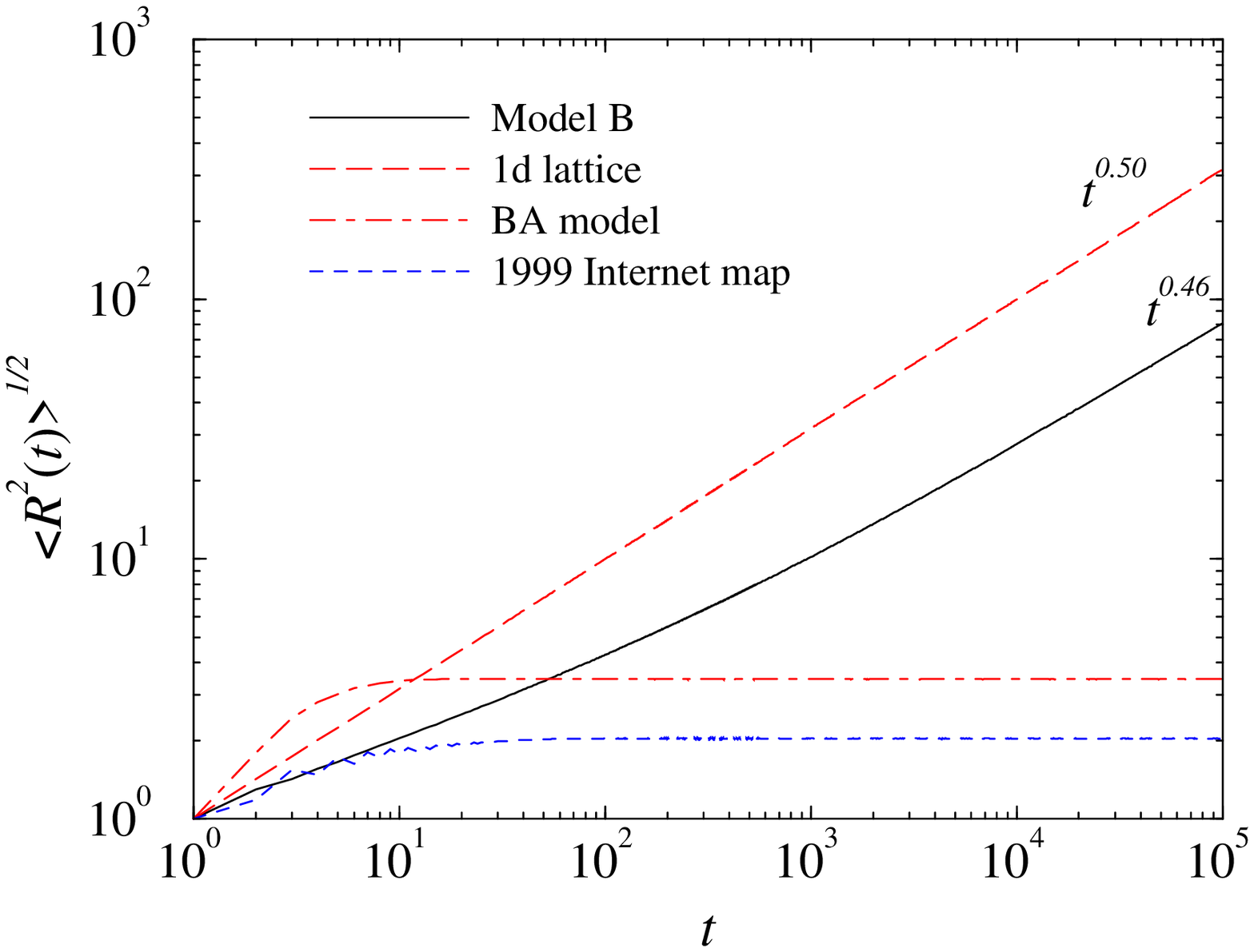,width=3in}}

\caption{Mean-square displacement of a random walker on the
  deactivation model with $m=3$, a one dimensional lattice, and the
  Barab{\'a}si-Albert model with $N=10^5$ nodes, as well as an Internet
  snapshot map from 1999 with $6301$ nodes.}
\label{fig:9}
\end{figure}

One of the most interesting issues related to SF networks is the
effects of their complex topological features on the dynamics of
spreading phenomena~\cite{pv01a,pv01b,lloyd01,moreno02} and the onset
of percolation transitions~\cite{barabasi00,newman00,havlin01}.  In
the case of random SF networks, where degree correlations are absent,
it has been found that the epidemic threshold is proportional to
$\avk/\fluck$ \cite{pv01a,pv01b}. Uncorrelated SF networks allow the
onset of large epidemics whatever the spreading rate of the infection.
This is a noticeable result that has a large impact in immunization as
well as control and design policies in real networks
\cite{psvpro,aidsbar}. On the other hand, most real networks show
non-trivial degree correlations and clustering properties as it is the
case in the present deactivation model. Similarly, the random removal
of nodes does not destroy the connectivity of SF networks with
$\gamma\leq 3$. In other words the percolation transition is absent,
and the networks are extremely robust to random damages
\cite{barabasi00,newman00,havlin01}.  A natural question is to know
whether or not the clustering properties of SF networks plus their
correlations alter the general results obtained for uncorrelated
networks. For this reason, several recent works have addressed the
effect of such correlations in the epidemic spreading occurring on
these networks \cite{structured,crucitti02,marian1,sander}.  In
particular, in Ref.~\cite{structured} it has been claimed the
existence of an epidemic threshold in the case of the deactivation
model for rule B.

The presence of a finite threshold in the deactivation model has been
traced back to the high clustering coefficient and the finite limit of
the average nearest neighbor connectivity of the largest
hubs~\cite{structured}.  On the other hand, we have shown here that
the average nearest neighbor connectivity in the system is diverging
with the system size.  What appears as more fundamental for the
properties of spreading in the deactivation model is its linear
structure, with a diameter that increases with $N$. In a coarse
grained picture, the epidemic spreading is dominated by the diffusion
of the disease on a linear chain. In order to
check this point, we have simulated a standard random walk in the B
model with $m=3$. In Fig.~\ref{fig:9} we plot the mean-square
displacement of the random walker, $\left< R^2(t) \right>^{1/2} =
\left< [r(t) - r(0)]^2 \right>^{1/2}$, where the brackets denote an
average over $250$ realizations of the random walk on $250$ different
networks. For a purely diffusive system, as would be the case of a one
dimensional lattice, we would expect a scaling $\left< R^2(t)
\right>^{1/2} \sim t^{1/2}$. For the deactivation model we observe a
slightly subdiffusive behavior, with a mean-square displacement
scaling as $\left< R^2(t) \right>^{1/2} \sim t^{0.46}$. We thus conclude
that dynamics on the deactivation model is almost purely diffusive, as
expected from its non small-world character. The analysis of spreading
and percolation properties in this network cannot therefore be
performed at the mean-field level \cite{pv01a,pv01b}, but must include
diffusion and most probably fluctuations, leading to a much more
complex formalism based in a field theory \cite{marro99}. For the sake
of comparison, we have also plotted in Fig.~\ref{fig:9} the
mean-square displacement of a random walker on a Barab{\'a}si-Albert
network \cite{barab99} and on a Internet snapshot map from 1999,
collected by the National Laboratory for Applied Network Research
(NLANR) \cite{nlanr}. As we observe, in these last two networks
$\left< R^2(t) \right>^{1/2}$ saturates very quickly to a constant
value, proportional to the network's diameter, indicating the presence
of a strong small-world component. The essential difference of the
diffusive properties between the Internet and the deactivation model
does not allow to extend the conclusions obtained from the model to
the spreading in the real system.

The same applies to percolation properties, that naturally exhibit a
finite threshold in this case.  The fact that spreading and
percolation properties on the deactivation model are similar to those
of regular lattices because of the absence of small-world features is
corroborated by the analysis of Ref.~\cite{crucitti02}, that shows how
the introduction of a small amount of shortcuts restores the usual
absence of a percolation threshold.  In this perspective, it would be
extremely interesting to have a detailed study of the epidemic
spreading properties in the case of the deactivation model with random
rewiring~\cite{klemm02b}, in order to assess the effect of clustering
and degree correlations in spreading processes in SF networks with
small-world properties.

\section{Acknowledgments}
This work has been partially supported by the European commission FET
Open project COSIN IST-2001-33555. R.P.-S. acknowledges financial
support from the Ministerio de Ciencia y Tecnolog{\'\i}a (Spain).

\end{document}